\documentclass{mem}
\usepackage{natbib}\usepackage{txfonts}\usepackage{balance}
\usepackage{graphicx}
\usepackage[a4paper]{hyperref}
\idline{75}{282}
\begin{document}
\def\teff{$T\rm_{eff }$}
\def\kms{$\mathrm {km s}^{-1}$}
\def\lsim{\,\lower2truept\hbox{${< \atop\hbox{\raise4truept\hbox{$\sim$}}}$}\,}
\def\gsim{\,\lower2truept\hbox{${> \atop\hbox{\raise4truept\hbox{$\sim$}}}$}\,}

\title{
Cosmic Radiative Feedback from Reionization
}

   \subtitle{}

\author{
R. \,Salvaterra\inst{1}, 
C. \, Burigana\inst{2,3},
R. \, Schneider\inst{4},
T. R. \, Choudhury\inst{5},
A. \, Ferrara\inst{6},
\and L. A. \, Popa\inst{7,2}
          }

  \offprints{R. Salvaterra}

\institute{ 
INAF, Osservatorio Astronomico di Brera, via E. Bianchi 46, I-23807 Merate
(LC), Italy. \email{salvaterra@mib.infn.it}
\and
INAF-IASF Bologna, 
Istituto di Astrofisica Spaziale e Fisica Cosmica di Bologna,
Istituto Nazionale di Astrofisica, via Gobetti 101, I-40129 Bologna, Italy 
\and
Dipartimento di Fisica, Universit\`a degli Studi di Ferrara,
via Saragat 1, I-44100 Ferrara, Italy
\and
INAF - Osservatorio Astrofisico di Arcetri, Largo Enrico Fermi 5, 
I-50125 Firenze, Italy
\and
Institute of Astronomy, Madingley Road, Cambridge CB3 0HA, UK
\and
SISSA/International School for Advanced Studies, Via Beirut 4, 
I-34100 Trieste, Italy
\and
Institute for Space Sciences, Bucharest-Magurele, Str. Atomostilor, 409, 
PoBox Mg-23, Ro-077125, Romania 
}

\authorrunning{Salvaterra et al.}

\titlerunning{Cosmic Radiative Feedback from Reionization}

\abstract{
We explore the effect of cosmic radiative feedback from the sources of
reionization on the thermal evolution of the intergalactic medium. We find 
that different prescriptions for this feedback predict quite different 
thermal and reionization histories. In spite of this, current
data can not discriminate among different reionization scenarios. We find
that future observations both from 21-cm and CMB experiments can be used to
break the degeneracy among model parameters provided that we will be able
to remove the foreground signal at the percent (or better) level. 
}
\maketitle{}

\section{Introduction}

It is well known that the temperature increase of the cosmic gas in ionized
regions leads to a dramatic suppression of the formation of low-mass 
galaxies (see Ciardi \& Ferrara 2005 for a review). We explore the impact of
this effect during cosmic reionization by  considering two different feedback
prescriptions: (i) {\bf suppression model} where galaxies can form stars
unimpeded provided that their circular velocity is larger than a critical
threshold, which is not fixed to a constant value but evolves according
to gas temperature \cite{CF06}; (ii) {\bf filtering model},
where, depending on the mass of the galaxy, the fraction of gas available
to star formation is reduced with respect of the universal value and it is
fully specified by the filtering mass at that redshift \cite{G00}.

\begin{figure}[t!]
\resizebox{\hsize}{!}{\includegraphics[clip=true]{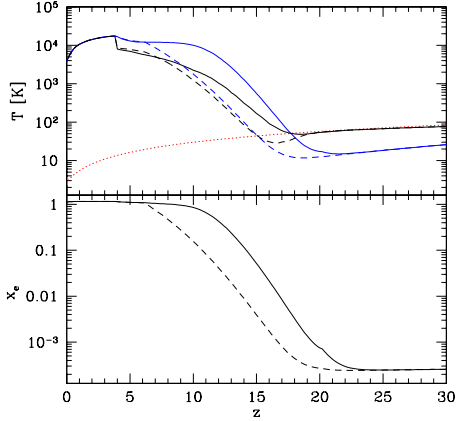}}
\caption{\footnotesize {\it Top panel:} redshift evolution of the spin 
(thick lines) and gas kinetic (thin lines) temperatures  predicted by the two 
models. Solid (dashed) lines refer to suppression (filtering) model. 
For comparison, we also show the evolution of the CMB temperature $T_\gamma$ (dotted line).
{\it Bottom panel:} corresponding evolution of the free electron fraction.
}
\label{history}
\end{figure}

We implement these two different radiative feedback prescriptions
into a physically-motivated and observationally tested model of reionization
\cite{CF05,CF06}. 
Although the two feedback prescriptions predicts quite different reionization 
and thermal histories (see Fig.~\ref{history}), in both scenarios it is possible to 
reproduce a wide range of observational data 
with a proper choice of few model parameters
(the redshift evolution of 
Lyman-limit absorption systems, the Gunn-Peterson and electron scattering
optical depths, the cosmic star formation history, and number counts in 
high-$z$ sources). Thus, we find that existing data are unable to discriminate
among the two reionization histories \cite{Schneider08}. We then
explore alternative methods to break these degeneracies using future
21-cm experiment \cite{Schneider08} and CMB anisotropy observations
\cite{Burigana08}.

\section{21-cm signal}

\begin{figure}[t!]
\resizebox{\hsize}{!}{\includegraphics[clip=true]{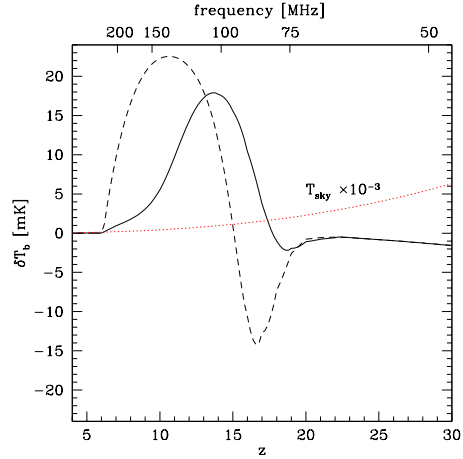}}
\caption{\footnotesize Predicted all-sky 21 cm brightness temperature relative to the CMB. Solid (dashed) line
refers to suppression (filtering) model. The upper x-axis shows a reference 
observational frequency scale. The dotted line shows an estimate of the foreground
signal at these frequencies, rescaled by a factor $10^{-3}$.
}
\label{21cm}
\end{figure}

\begin{figure}[ht!]
\resizebox{\hsize}{!}{\includegraphics[clip=true]{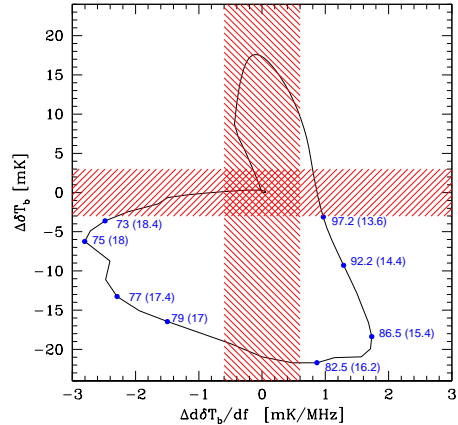}}
\caption{\footnotesize 
Difference (filtering~$-$~suppression)
in the brightness temperature and gradients between the two models (see text). The
shaded areas represent the regions where the signals will not be distinguishable given the 
foreseen sensitivity of future 21 cm experiments. Numbers along the curves indicate the 
frequency in MHz (and the corresponding redshift) at which the signals would be observed.}
\label{fig:obs}
\end{figure}

The spin temperature $T_S$, 
which represents the excitation temperature of
the 21 cm transition,
determines whether the signal will appear in emission
(if $T_S > T_\gamma$) or in absorption (if $T_S < T_\gamma$).
Given the different gas ionization fraction and spin temperature evolution in the range 
$7 \lsim z \lsim 20$,  the two models predict different global 21 cm background signals in the observed
frequency range $75 \mathrm{MHz} \lsim \nu \lsim 200$~MHz. The largest differences in the two models
are represented by a $\sim 15$~mK absorption feature in the range 75-100 MHz in filtering model
(which is nearly absent in suppression model), and by a global shift of the emission feature preceding
reionization towards larger frequencies in the same model, 
as shown in Fig.~\ref{21cm} (see Schneider et al. 2008 for all details). 
Since the frequency dependence of the signals and foregrounds are different,
the gradient of the brightness fluctuation with frequency 
might help to discriminate the signal from the
relatively smooth foreground in the comparison between the two models, 
as shown in Fig.~\ref{fig:obs}.
Single dish observations with
existing or forthcoming low-frequency radio telescopes such as LOFAR, 21CMA, MWA, and SKA 
can achieve mK sensitivity allowing the identification of these signals provided that foregrounds, 
which are expected to be three orders of magnitude larger, can be accurately subtracted.
The best observational frequencies to discriminate the radiative feedback models through their 21 cm
background signal are 73-79 MHz and 82.5-97.2 MHz, where the expected differences in brightness 
temperatures and gradients are large enough to be detectable with future 21 cm experiments.

\section{CMB anisotropies}

\begin{figure}[t!]
\resizebox{\hsize}{!}{\includegraphics[clip=true,angle=90]{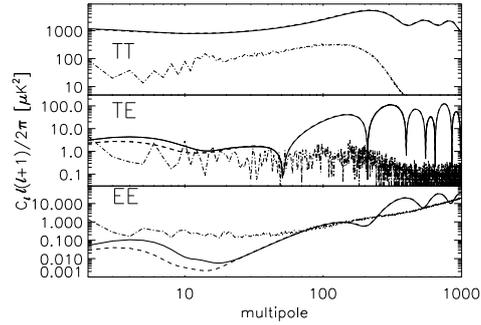}}
\caption{\footnotesize
APS of CMB anisotropies
for the three considered modes TT, TE, EE, reported in each panel
(solid lines: suppression model; dashes: filtering model).
Thick lines denote correlation, while thin lines denote anticorrelation
(appearing for the TE mode at $\ell \gsim 50$).
The APS of the foreground, dominated by the Galactic contribution,
is reported for comparison (dot-dashes).
See also the text. [Results expressed in terms of thermodynamic temperature fluctuations].}
\label{cmb_cl}
\end{figure}

In this section, we discuss how to discriminate between the two radiative
feedback prescriptions (and corresponding reionization histories) with future CMB
data (see Burigana et al. 2008 for all details). 

By exploiting the ionization and kinetic temperature histories shown in
Fig.~\ref{history} we compute the Comptonization and the 
free-free\footnote{The values found for $y_B$ 
should be considered as lower limits, since it is computed in the diffuse ``averaged density'' 
approximation; so, a correction factor $\, \approx \, <n^2>/<n>^2 \; \; > 1$, coming from a 
proper inclusion of the treatment of density contrast in the intergalactic medium, should 
be applied.} 
distortion parameters. We find
$u \simeq 1.69\times10^{-7}$, $y_B \simeq 9.01\times10^{-10}$ and
$u \simeq 9.65\times10^{-8}$, $y_B \simeq 5.24\times10^{-10}$ respectively
for the suppression and the filtering model: these values are clearly well
below the COBE/FIRAS limits.
The two models show similar ionization and thermal histories at $z \lsim 6$
while important differences are predicted at $z \gsim 6$. These explain the
different spectral distortion levels generated in the two cases.
The expected Comptonization distortions are comparable to
those that could be in principle observed by a future generation of
CMB spectrum experiments.

We then compute the angular power spectrum (APS) of CMB anisotropy including the ionization
history in a dedicated Boltzmann code.
Our results
are reported in Fig.~\ref{cmb_cl}. Having neglected for simplicity
tensor perturbations, we focus here on the TT (total intensity, i.e. temperature),
TE (temperature-polarization cross-correlation), and EE polarization modes of
the CMB anisotropy APS.
We display also the APS of the foreground
in the V band (centred at 61 GHz) of WMAP 3yr
data\footnote{http://lambda.gsfc.nasa.gov/product/map/current/},
a frequency where the foreground is found to be minimum
(or almost minimum) in both temperature and polarization
at the angular scales larger than $\sim 1^\circ$
of relevance in this context.

\begin{figure}[t!]
\resizebox{\hsize}{!}{\includegraphics[clip=true,angle=90]{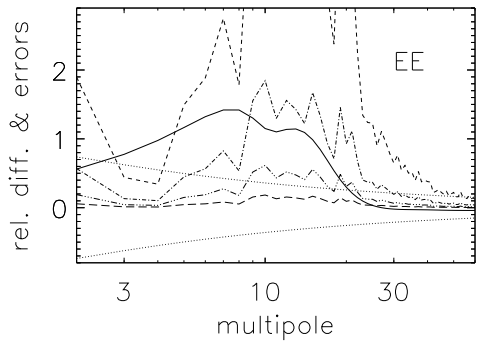}}
\caption{\footnotesize
Relative difference
between the (EE mode) APS of CMB anisotropies for the suppression and filtering models (thick solid lines) is
compared with the cosmic and sampling variance limitation (dotted lines) and
different foreground residual (dot-dashed lines). See the text for more 
details.}
\label{cmb}
\end{figure}

Fig.~\ref{cmb}  shows the relative difference
between the EE mode APS of CMB anisotropies for the suppression and filtering models
reported in Fig.~\ref{cmb_cl} 
compared with the cosmic and sampling variance limitation
corresponding to a sky coverage of $\simeq 74$ per cent
(region between the dotted lines).
We report also for comparison the APS from a potential residual foreground
(dot-dashes)
corresponding to different values of the relative accuracy (at APS level) of the component 
separation method in the considered range of multipoles: 
0.1 (dashed line), 0.03 (dots-dashed line), 
0.01 (three dots-dashed line), and 0.003 (long dashes).
The difference between the two considered models is significantly larger 
than the cosmic and sampling variance over a interesting range of multipoles
($\ell \sim 5-15$). The main limitation derives from a possible
residual foreground contamination: as evident, a foreground removal 
accuracy at per cent level (or better)
in terms of APS is necessary to accurately exploit the 
information contained in CMB polarization about the cosmological reionization process. 
This calls for a further progress in component separation techniques in polarization
and for an accurate mapping of the (mainly Galactic) polarized foregrounds 
in radio
and far-IR bands
to complement microwave surveys by improving
both the foreground physical modeling 
and the component separation accuracy.

While the difference between the APS of CMB 
anisotropies in the two models is overwhelmed by (respec. not significantly larger than)
the cosmic and sampling variance for the TT (respec. the TE)
mode, we find that future accurate large sky coverage observations of EE polarization mode 
(e.g. from the forthcoming ESA {\it Planck} 
mission\footnote{http://www.rssd.esa.int/planck}
or from the next generation of polarization dedicated satellites, as CMBPol 
and B-Pol\footnote{http://www.th.u-psud.fr/bpol/index.php}) 
can be used to discriminate between the two reionization histories.

\section{Conclusions}

We have explored the effect of cosmic radiative feedback from the sources of
reionization on the thermal evolution of the intergalactic medium. We 
implemented two different prescriptions for this feedback into a well-tested,
physically-motived model of the early Universe. We found 
that different prescriptions for this feedback predict quite different 
thermal and reionization histories. In spite of this differences, current
data can not discriminate among different reionization scenarios. Therefore,
we explored alternative methods to break this degeneracies using 21-cm
experiment and CMB anisotropy observations. We found that future data can 
distinguish among different reionization histories provided that we would be 
able to remove the foreground signal at the percent (or better) level. 

\bibliographystyle{aa}

\end{document}